\documentclass[11pt]{article}
\usepackage{sint}
\usepackage{macros}
\usepackage[dvips]{epsfig}
\begin{document}
\begin{titlepage}
\begin{flushleft}
   May 2001
\end{flushleft}
\vspace*{-2cm}
\begin{flushright}
   Bicocca-FT-01-12\\
   CERN-TH/2001-120\\
   HU-EP-01/16\\
   MS-TP-01-2\\
   DESY 01-052\\
   MPI-PhT/2001-12\\
   LTH 501\\
\end{flushright}
\vskip 0.5 cm
\begin{center}
 {\Large\bf First results on the running coupling in QCD
            with two massless flavours}
\end{center}
\vskip 0.5 cm
\centerline{
\includegraphics[width=2.5cm]{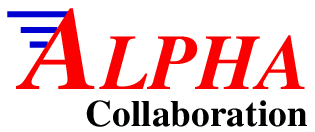}
}
\begin{center}
{\large Achim~Bode$^{\scriptscriptstyle a}$,
        Roberto~Frezzotti$^{\scriptscriptstyle b,c}$,
        Bernd~Gehrmann$^{\scriptscriptstyle d}$,
        Martin~Hasenbusch$^{\scriptscriptstyle d}$,
        Jochen~Heitger$^{\scriptscriptstyle e}$,
        Karl~Jansen$^{\scriptscriptstyle f}$,
        Stefan~Kurth$^{\scriptscriptstyle d}$,
        Juri~Rolf$^{\scriptscriptstyle d}$,
        Hubert~Simma$^{\scriptscriptstyle g}$,
        Stefan~Sint$^{\scriptscriptstyle c,h}$,
        Rainer~Sommer$^{\scriptscriptstyle g}$,
        Peter~Weisz$^{\scriptscriptstyle i}$,
        Hartmut~Wittig$^{\scriptscriptstyle j}$ and
        Ulli~Wolff$^{\scriptscriptstyle d}$}
\end{center}
\vskip 2.3ex
\begin{flushleft}
$^{\scriptstyle a}$ CSIT, Tallahassee, USA\\
$^{\scriptstyle b}$ Dipt. di Fisica, Univ. di Milano Bicocca,
Milano, Italy\\
$^{\scriptstyle c}$ CERN-TH, Geneva, Switzerland\\
$^{\scriptstyle d}$ Institut f\"ur Physik, Humboldt Universit\"at, 
Berlin, Germany\\
$^{\scriptstyle e}$ Institut f\"ur Theor. Physik, Universit\"at M\"unster,
M\"unster, Germany\\
$^{\scriptstyle f}$ NIC, Zeuthen, Germany\\
$^{\scriptstyle g}$ DESY, Zeuthen, Germany\\
$^{\scriptstyle h}$ Dipt. di Fisica, Univ. di Roma, Tor Vergata, 
Rome, Italy\\ 
$^{\scriptstyle i}$ Max-Planck-Institut f\"ur Physik, M\"unchen, Germany\\
$^{\scriptstyle j}$ Dept. of Mathematical Sciences,
Univ. of Liverpool, Liverpool, UK
\end{flushleft}
 {\bf Abstract}
We report on the non-perturbative computation of the running coupling 
of two-flavour QCD in the 
Schr\"odinger functional scheme.  The corresponding
$\Lambda$-parameter, which  
describes the coupling strength
at high energy, is 
related to a low energy scale which still
remains to be connected
to a hadronic ``experimentally'' observable
quantity.
We find the {\em non-perturbative} evolution of the coupling
important to eliminate a significant contribution to the
total error in the estimated
$\Lambda$-parameter. 
\\
{\it Keywords:} lattice QCD; dynamical fermions; running coupling; 
renormalization.\\
{\it PACS:} 11.15.Ha; 12.38.Gc; 12.38.Bx; 11.10.Gh; 11.10.Hi
\vskip 0.7ex
\newpage

\vfill

\eject

\end{titlepage}


\section{Introduction}

Under lattice regularization predictions of
renormalized quantum field theories
emerge as universal properties of critical points of
models in the appropriate universality class.
In this way the theory is defined independently of perturbation theory
and may for instance be evaluated numerically.
Predictive power resides in a surplus of relations between observables
over free parameters in the model, and it becomes a well-defined
question which part of these relations is
amenable to approximation by renormalized perturbation theory.
In QCD the standard expectation is that quantities associated with energies
large compared to typical hadron masses can be perturbatively related
to each other. 
If one is limited to this calculational framework, a small number of
input parameters associated with large normalization energy $\mu$,
like the coupling $\alphas (\mu)$
and the quark masses for each flavour $\mqf (\mu)$, have to be determined
from experiment and then lead to many successful predictions of
perturbative QCD.

By lattice techniques it becomes possible to look beyond the perturbative
horizon. Consequently a lot of activity goes and went into extracting
information on the hadronic low energy sector. In particular the free
parameters are determined in this case by inputting quantities like
some hadron masses or $\Fpi$. Then the high energy sector can in principle
be predicted by {\em evaluating} 
$\alphas (\mu)$ and the quark masses
for
$\mu \gg \Fpi$.
Such calculations relating different orders of magnitude
of physical scales represent a formidable numerical problem.
Beside the dissimilar physical scales, infrared and ultraviolet cutoffs
have to be extrapolated to their respective limits.
A number of such direct approaches have nevertheless been tried,
and some have found their entry into the particle data table
\cite{PDG-AIDA} 
as 
one of the most accurate determinations of $\alpha(M_{\rm Z})$.
In view of the very limited parameter range accessible to simulation 
we find it difficult to be confident about the systematic errors of these
determinations.
A computation requiring more steps but also offering much more
control of systematic errors becomes feasible
by the recursive finite size method using the Schr\"odinger functional.
This technique has been
developed by our
collaboration over the last years and is reviewed in \cite{revRS} and
\cite{revML}.

In this publication we present first numerical results toward the extension of
the method from the zero flavour (quenched) approximation to QCD with
two light flavours which are taken massless.
Sect.~2 summarizes the most essential results of our previous work in the
present context. In sect.~3 
the computational strategy for the $\Lambda$-parameter characterizing
the coupling at large energy is outlined, followed by numerical results
in sect.~4 and some conclusions.

\section{\SF setup}

To connect hadronic and perturbative scales in QCD
an intermediate renormalization scheme has been devised
where the finite system size $L$ is used as a renormalization
scale. More precisely, we consider the \SF
given by the partition function of QCD on
a cylinder of size $T\times L^3$ in euclidean space
\be
\re^{-\Gamma} = \int_{T\times L^3} \rD[U,\psi,\psibar]\, \re^{-S}.
\ee
In the lattice regularized form 
we integrate over SU(3) gauge fields $U$ with the Wilson action
and two flavours of O($a$) improved Wilson quarks $\psi, \psibar$.
Boundary conditions in the spatial directions of length $L$
are periodic \cite{LNWW} for $U$
and periodic up to a global phase \cite{SS} 
$\theta=\pi/5$ for
$\psi,\psibar$.
In Euclidean time Dirichlet boundary conditions are imposed at $x_0=0,T$ by
fixing spatial links to diagonal SU(3) matrices
that are precisely specified in terms of $L$ and two angles $\eta$ and $\nu$
(point `A' in \cite{SFSU3}), and we also take $T=L$. 
The quark fields on the boundary surfaces
\cite{LSSW}
are given by Grassmann values $\rho,\rhobar$ and $\rho',\rhobar'$,
which are used as sources that are set to zero after differentiation.

To achieve the convergence to the continuum limit 
at a rate proportional to the squared lattice spacing
$a^2$ a number of irrelevant
operators have to be tuned. The coefficient $\csw$ 
of the clover term \cite{SW}
is set to the non-perturbative
values quoted in parameterized form in \cite{JS}.
In the \SF 
at vanishing quark mass, that we consider here, 
the coefficients $\ct$ and $\cttilde$ of
two additional boundary counter terms \cite{LSSW}
have to be adjusted. Here we have to content ourselves with
perturbative estimates at one and two-loop accuracy \cite{LW,BWW}.

Since for Wilson fermions chiral symmetry only emerges in the continuum
limit, the bare mass parameter is additively renormalized. 
For this reason we trade it for a quark mass
defined by the 
PCAC relation evaluated using suitable states
\cite{LSSWW}. Let $\fA(x_0)$ and $\fP(x_0)$ be the matrix
elements of the axial current and the pseudoscalar density defined in (2.1)
and (2.2) of \cite{LSSWW} with the gluonic
boundary fields assuming the values quoted above. 
We form
the ratio
\be
m(x_0) = \frac{\frac12 (\del_0^{\phantom{*}} +
\del_0^*)  \fA(x_0)+\cA a \del_0^{\phantom{*}} 
\del_0^*\fP(x_0)}{2\fP(x_0)}
\ee
with forward (backward) derivative 
$\del_0^{\phantom{*}}$ ($\del_0^*$).
For the current improvement coefficient $\cA$ its one-loop value
\cite{LW} is taken. We now 
define the bare current mass
\be
m_1 = \left\{\begin{tabular}{c@{ for }l}
$m(T/2)$ & even $T/a$\\
$[m((T-a)/2)+m((T+a)/2)]/2$ & odd $T/a$.
\end{tabular}\right.
\ee
An alternative definition $m_2$ just differs by averaging 
$m(x_0)$ and $m'(x_0)$, where the latter is defined \cite{LSSWW} with the
sources $\rho',\rhobar'$ at the $x_0=T$ boundary
leading to $\fA',\fP'$. These masses are expected to differ at O($a^2$).
With either of them vanishing, the chirally symmetric continuum limit
may be approached.

The coupling $\gbar^2$ and the additional universal dimensionless
observable $\vbar$ are related to the \SF by
\be
\frac{\del \Gamma}{\del \eta} = k \left\{
\frac{1}{\gbar^2(L)} - \nu \vbar(L)
\right\},
\label{gbardef}
\ee
where $k$ is a known \cite{SFSU3} normalization fixed by demanding
$\gbar^2=g_0^2+\rO (g_0^4)$ with the bare coupling $g_0$.

\section{Computational strategy for the $\Lambda$-parameter}

Our method to extract $\Lambda$, which characterizes
the behaviour of $\gbar^2$ at asymptotically large energy,
follows the strategy
used in \cite{CLSW}. By continuum extrapolation we construct
the non-perturbative step scaling function (SSF)
\be
\sigma(u) = \gbar^2 (2L)\bigr|_{\gbar^2 (L)=u,m_1=0}
\label{defsigma}
\ee
for a number of $u$-values such that  
by interpolation we control it over the range that will be needed.
Then a value $\umax$ is selected (initially by guesswork)
such that the associated scale $\Lmax$
where $\gbar^2(\Lmax)=\umax$ is in the hadronic range.
By recursively solving $n$ times
\be
\sigma(\gbar^2 (L/2)) =  \gbar^2 (L)
\ee
starting with $L=\Lmax$ we obtain values for $\gbar^2(2^{-n}\Lmax)$.
Finally, for sufficiently large $n$, this coupling is perturbative
and we use
\bea
\Lambda \Lmax &=&  2^n (b_0\gbar^2)^{-b_1/2b_0^2} 
\exp\left\{ -\frac1{2b_0\gbar^2}  \right\} \nonumber\\
&& \times
\exp\left\{ -\int_0^{\gbar} dx \left[        
\frac1{\beta(x)} + \frac1{b_0 x^3} -\frac{b_1}{b_0^2 x}  \right]\right\}\
\label{Lambdapert}
\eea
to derive $\Lambda$ in terms of $\Lmax$. Here the three-loop 
$\beta$-function\footnote{
We now use $b_2=0.06/(4\pi)^3$ given in the second erratum to \cite{BWW}
which has become necessary due to the revision of \cite{Haris}
in May 2001. Since our analysis depends on these results,
an independent check seems desirable. It will be partially supplied
in the near future \cite{AH}.
}
for the SF-scheme with two flavours 
\cite{BWW} is used and $b_0,b_1$ are its universal coefficients.
On the right-hand side  $\gbar^2$ is understood to be inserted
at the scale
$2^{-n}\Lmax$.
The admissibility and accuracy of renormalized perturbation theory
can be probed by checking the stability of the result when varying $n$.

In a later series of simulations we shall have to relate $\Lmax$
to a truly physical scale, for instance by computing
$\Lmax F_\pi$. 
The guess for $\umax$ will be confirmed then, if a number of order one
is found, i.~e. the multiple scale problem is avoided.
The relation between $\Lambda$, which corresponds to the
SF-scheme, and the $\MSbar$-scheme is given by \cite{SS,BWW}
\be
\Lambda_{\MSbar}=2.382035(3) \Lambda.
\ee

\section{Numerical results}
The continuum SSF 
is given by the limit
\be
\sigma(u)=\lim_{a/L \to 0} \Sigma(u,a/L),
\ee
where $\Sigma$ is defined like $\sigma$ in eq.~(\ref{defsigma})
but interpreted at finite resolution $a/L$.
Since PCAC becomes an operator relation in the continuum limit only,
we adopt the convention to always tune $m_1(L/a)$ to zero on the small
lattice. The corresponding value $m_1(2L/a)$ measured at resolution $a/2L$
is expected to differ by O($a^2$) from $m_1(L/a)$. In the same way
we expect $m_1(L/a)-m_2(L/a)=\rO(a^2)$ for our alternative
definition $m_2$ of the bare current quark mass.
We tested these expectations on our data and compared with one-loop
perturbation theory in Fig.~1 for several couplings.
\begin{figure}[ht]\label{d_and_e}
  \begin{center}
    \includegraphics[width=12cm]{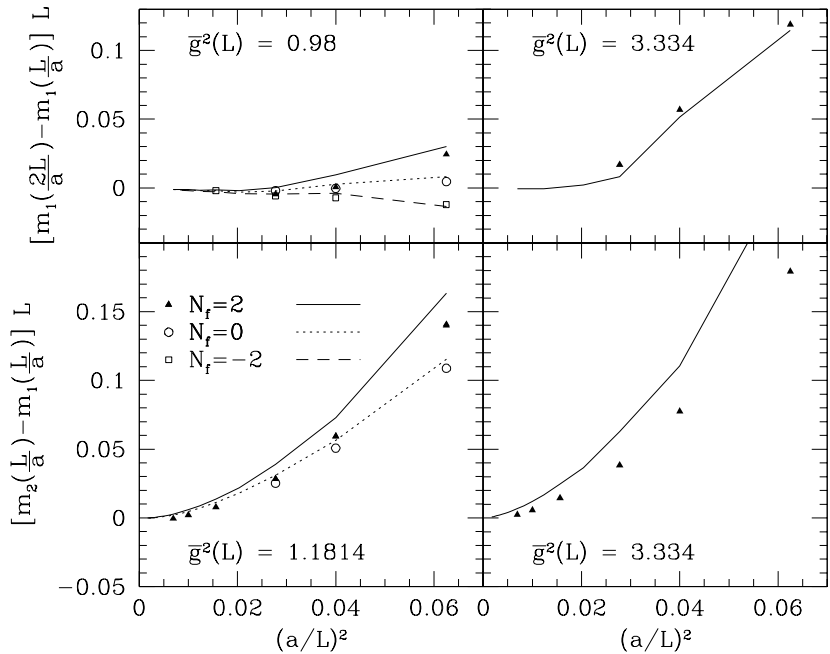}
  \end{center}
\caption{Lattice artefacts of PCAC masses. Perturbative results for integer
$L/a$ are connected by lines.}
\end{figure}
Where available we include
together with our present $N_{\rm f}=2$ data also quenched and 
bermion \cite{GJKW}
($N_{\rm f}=-2$) results. We conclude that lattice artefacts behave
non-pathologically and similar to perturbative expectations.
For the accessible 
range of resolutions they
happen to be dominated by terms of higher order than
the expected $a^2$-contributions.

Another place to study the approach to universal continuum
behaviour is the relation between
$\vbar$ and $\gbar$ defined in (\ref{gbardef}),
\be
\vbar = \omega(\gbar^2) = \lim_{a/L\to 0} \Omega(\gbar^2,a/L) .
\ee
In perturbation theory $\Omega$ is known to two-loop order,
\be
\Omega(u,a/L) = (v_1+v_2 u)\, \Bigl[1+\epsilon_1(a/L)+\epsilon_2(a/L) u\Bigr] 
+ \rO(u^2),
\label{vbarcorr}
\ee
and $\epsilon_1,\epsilon_2$ encode
the perturbative artefacts. 
\begin{figure}[ht]\label{vbar_nf2a}
  \begin{center}
    \includegraphics[width=7cm]{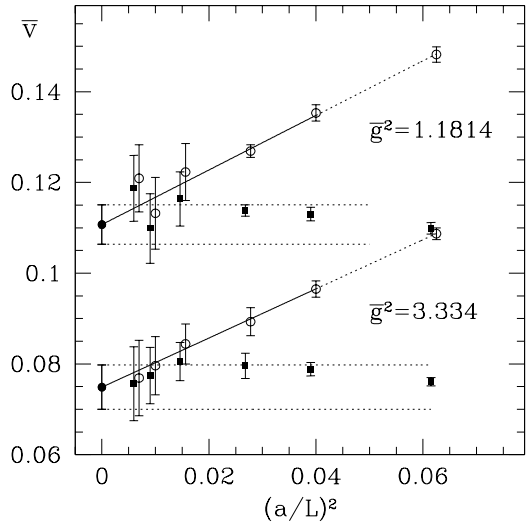}
  \end{center}
\caption{Circular symbols are data for $\vbar$
including a continuum extrapolation. Square symbols
have perturbative lattice artefacts cancelled.
}
\end{figure}
In Fig.~2 data for two different values of the
coupling are plotted. The square symbols refer to the improved
observable, where the Monte Carlo results have been divided by
the perturbative lattice artefacts $1+\epsilon_1(a/L)+\epsilon_2(a/L) u$ 
as first discussed
in \cite{SU2}. We conclude that a much smoother continuum limit
is achieved in this way.

At present, due to limitations of computing power, we have
results with sufficient
statistics only for $\gbar^2$ on lattices with $L/a \le 12$~. 
The resulting values of $\Sigma$ are collected in table~1.
The algorithmic aspects of these simulations have been presented in \cite{ALGS}.
\begin{table}[htbp]
\begin{center}
  \begin{tabular}{cll|ll} \hline
    $L/a$ & \multicolumn{1}{c}{$u$}  & \multicolumn{1}{c}{$\Sigma$} 
          & \multicolumn{1}{|c}{$u$}  & \multicolumn{1}{c}{$\Sigma$} \\ \hline
    4     & $0.9793(7)$   & $1.0643(34) \rule{0ex}{1ex}$
          & $1.5031(12)$   & $1.720(5)  $  \\
    5     & $0.9793(6)$   & $1.0721(39) $
          & $1.5033(26)$   & $1.737(10) $  \\
    6     & $0.9793(11)$  & $1.0802(44) $
          & $1.5031(30)$   & $1.730(12) $  \\[0.5ex]
    4     & $1.1814(5)$   & $1.3154(55) $
          & $2.0142(24)$   & $2.481(17) $  \\
    5     & $1.1807(12)$   & $1.3287(59)$
          & $2.0142(44)$   & $2.438(19) $  \\
    6     & $1.1814(15)$   & $1.3253(67)$
          & $2.0146(56)$   & $2.508(26) $  \\[0.5ex]
    4     & $1.5031(10)$   & $1.731(6)  $ 
          & $2.4792(34)$   & $3.251(28) $  \\
    5     & $1.5031(20)$   & $1.758(11) $
          & $2.4792(73)$   & $3.336(50) $  \\
    6     & $1.5031(25)$   & $1.745(12) $
          & $2.4792(82)$   & $3.156(55) $  \\[0.5ex]
    4     & $1.7319(11)$   & $2.058(7)  $
          & $3.334(11)$    & $5.298(85) $  \\
    5     & $1.7333(32)$   & $2.086(21) $
          & $3.334(15)$    & $5.41(12)  $  \\
    6     & $1.7319(34)$   & $2.058(20) $
          & $3.326(20)$    & $5.68(13)  $  \\[0.5ex] \hline
\end{tabular}
\end{center}
\caption[Table]{Data for the lattice step scaling function 
$\Sigma(u,a/L)$. For the left part $\ct (g_0)$ was set
to its one-loop value, whereas the two rightmost columns
have been obtained with the two-loop result \cite{BWW}.}
\end{table}

To estimate a continuum value for $\sigma$ from lattices with $L/a=4,5,6$
(together with the lattices at the doubled lengths)
we adopt the following procedure. First we perturbatively correct
the data with a factor analogous to the one in eq.~(\ref{vbarcorr})
\be
\Sigma(u,a/L) \rightarrow \Sigmaimpr(u,a/L)=
\frac{\Sigma(u,a/L)}{1+\delta_1(a/L)u+\delta_2(a/L)u^2}
\label{Sigmaimpr}
\ee
with the series for the artefacts known up to two-loop order.
They depend, of course, on the details of the action chosen.
As the two-loop boundary improvement coefficient $\ct (g_0)$ became available
only during our simulations they were partly carried out with its
one-loop value (left part of table~1) and only later with the two-loop
value. Hence two different sets of $\delta_2$ had to be 
used in (\ref{Sigmaimpr}). 
We found the values of $\Sigma^{(2)}(u,a/L)$
for $L/a=5,6$ constant within errors and fitted them to a constant
(i.~e. just combined them)
as our present continuum estimates.
They are found in table~2 together with 
$\Sigma^{(2)}$ at resolution $L/a=4$ for the estimation of systematic errors
(see below).
\begin{table}[htbp]
\begin{center}
\begin{tabular}
{lll}
\hline
\multicolumn{1}{c}{$u$} & \multicolumn{1}{c}{\rule{0ex}{2.5ex} $\sigma(u)$}
                        & \multicolumn{1}{c}{$\Sigmaimpr(u,1/4)$} \\ 
\hline
 0.9793 & 1.0768(30) & 1.0686(35) \\
 1.1814 & 1.3277(46) & 1.3199(55) \\
 1.5031 & 1.7489(85) & 1.7332(60) \\
 1.7319 & 2.063(15)  & 2.0562(72) \\
 1.5031 & 1.750(8)  & 1.7477(56) \\
 2.0142 & 2.494(16)  & 2.535(18)  \\
 2.4792 & 3.304(38)  & 3.338(28)  \\
 3.3340 & 5.65(10)   & 5.491(90)  \\
\hline
\end{tabular}
\end{center}
\caption[Table]{Numerical results for $\sigma(u)$.}
\end{table}
In Fig.~3 the analogous procedure can be judged in the quenched case, where
many more data are available.
The averages of  the points at $L/a=5,6$ in each series lead to the
dotted lines and are to be compared with the full extrapolation (points
at $a/L=0$).

\begin{figure}
  \begin{center}
    \includegraphics[width=10cm]{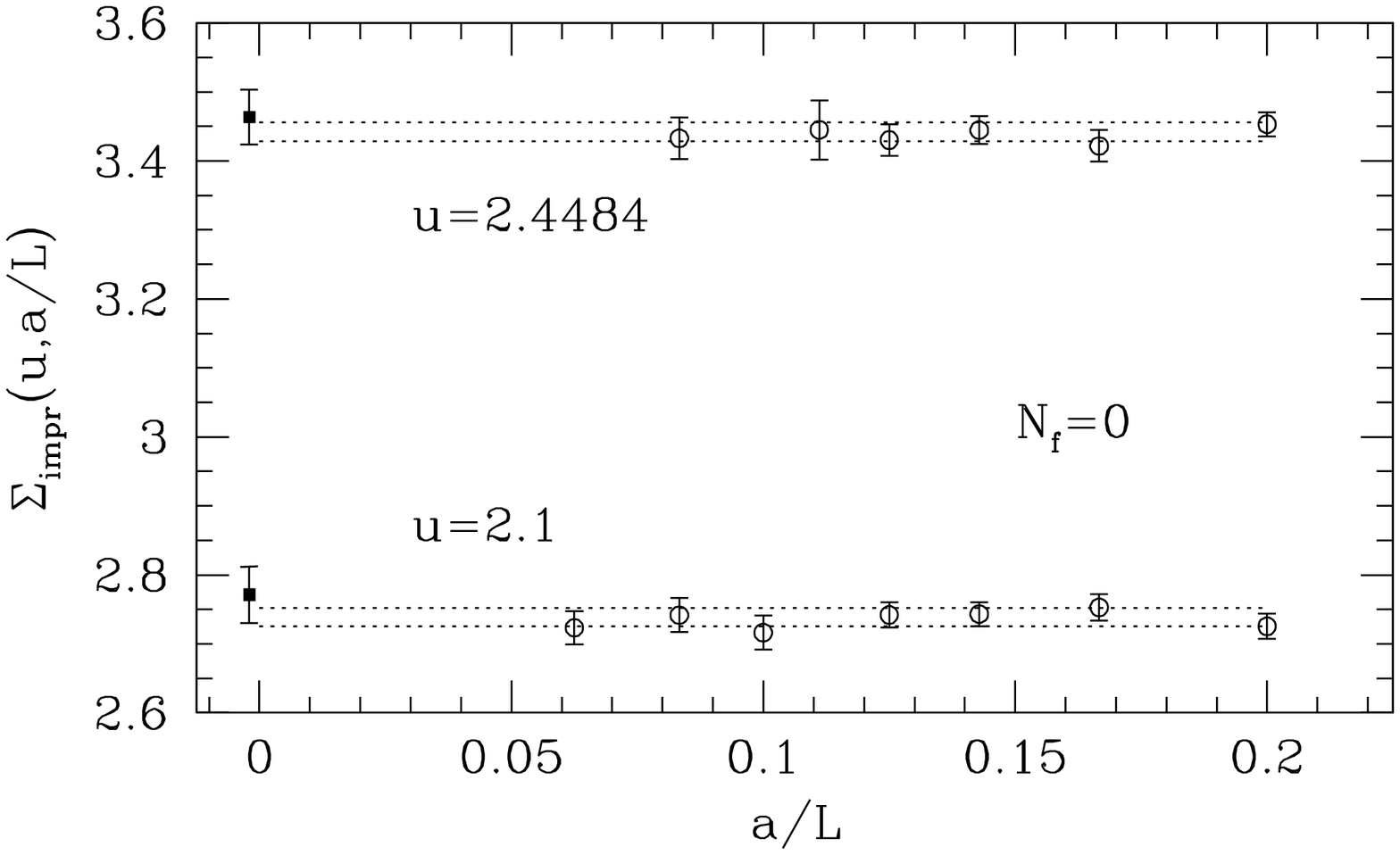}
  \end{center}
\caption{Quenched results for $\Sigmaimpr$ to illustrate
our extrapolation procedure. Data are mostly from the literature 
[9,16]
apart from the two finest resolutions at $u=2.1$ which were
obtained on APEmille at Zeuthen.
}
\end{figure}

We interpolate the values of table~2 by fitting $\sigma(u)$ to a sixth 
order polynomial with the first three coefficients constrained to their
perturbative values.
The resulting SSF is shown in Fig.~4.
It differs from the quenched SSF by an amount predicted well by perturbation
theory for weak 
coupling. For values above about 2.5 the three-loop term 
contributes significantly to
the $\beta$-function but actually enhances  
the growing gap  between Monte Carlo results and 
perturbation theory.
\begin{figure}
  \begin{center}
    \includegraphics[width=8cm]{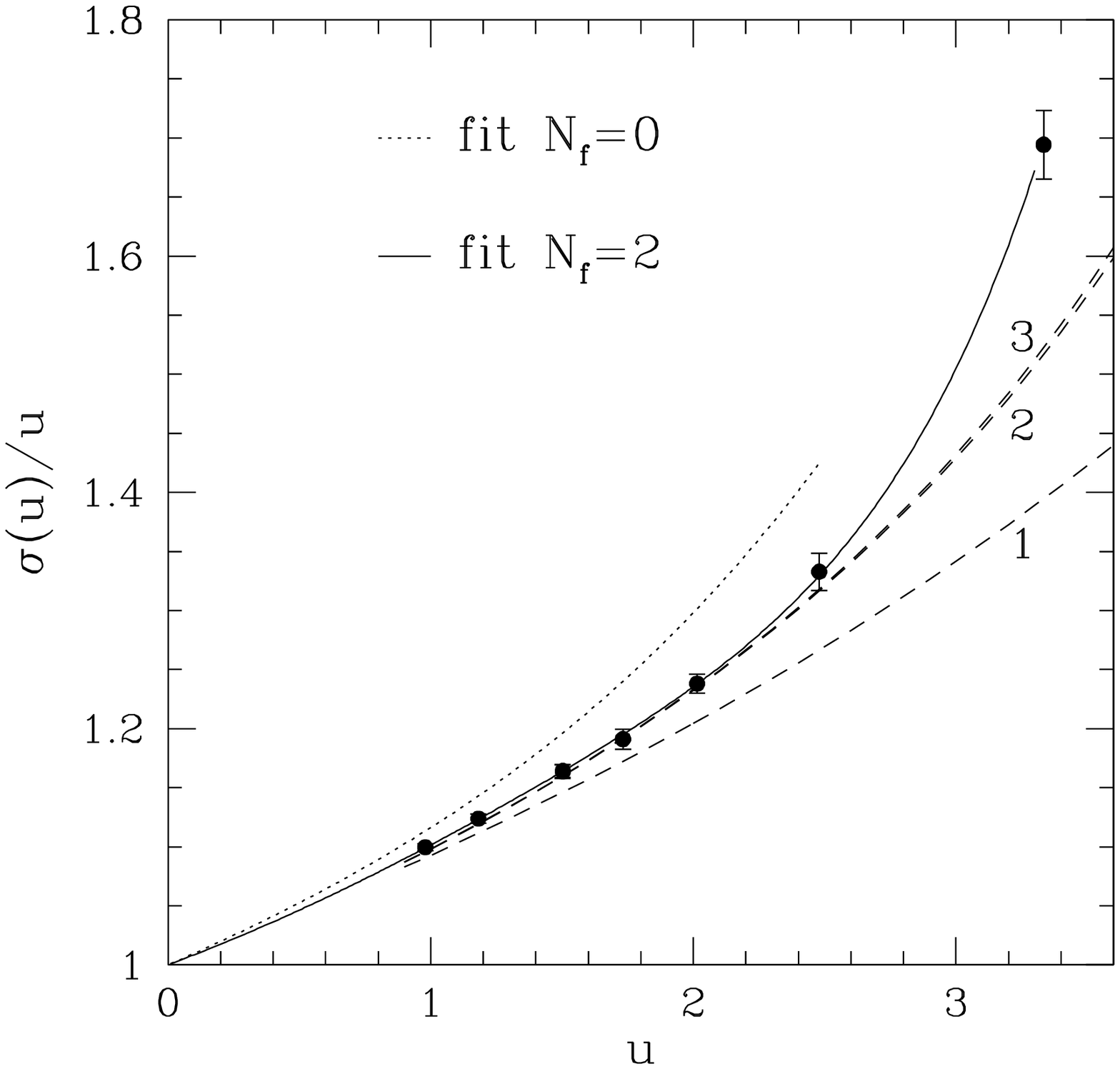}
  \end{center}
\caption{Step scaling function for $N_{\rm f}=2$ and $N_{\rm f}=0$ 
for comparison. Dashed lines are perturbative results from integrations
with the one-loop and, 
hardly distinguishable, with the two- and three-loop $\beta$-function.
}
\end{figure}

The fitted form for $\sigma$ is employed to estimate $\Lambda\Lmax$ 
in the way described in sect.~2 starting from $\umax=3.3$ and from $\umax=5$.
Statistical errors are obtained by propagating the errors
of the primary Monte Carlo data through the whole analysis,
and the inclusion of another parameter in the interpolating fit
for $\sigma(u)$
gave only negligible changes.
In this way we find the numbers in table~3.
\begin{table}[htbp]
\begin{center}
\begin{tabular}
{ccc|ccc}
\hline
  & \multicolumn{2}{c|}{$\gbar^2(\Lmax)=3.3$} & 
  & \multicolumn{2}{c}{$\gbar^2(\Lmax')=5$} \\
$n$ & continuum & $L/a=4$ & $n$ & continuum & $L/a=4$ \\ 
\hline
5 & 1.84(4) & 1.80 & 6 & 1.24(5) & 1.19 \\
6 & 1.86(5) & 1.78 & 7 & 1.26(5) & 1.17 \\
7 & 1.88(6) & 1.76 & 8 & 1.28(6) & 1.15 \\
\hline
\end{tabular}
\end{center}
\caption[Table]{Values estimated for $-\ln(\Lambda\Lmax)$
for two examples of $\Lmax$.}
\end{table}
In the  columns labelled by $L/a=4$ we have replaced
our continuum estimates $\sigma(u)$ by $\Sigmaimpr(u,1/4)$.
We regard the difference between the two columns as our present systematic
uncertainty\footnote{A somewhat smaller value
would be obtained if we took the magnitude of our
perturbative improvement for lattice artefacts,  
$\Sigmaimpr(u,1/5)-\Sigma(u,1/5)$, as an estimate of the systematic error.}
 and quote at the moment
\bea
\ln(\Lambda\Lmax) &=& -1.9(2) \quad [\gbar^2(\Lmax)=3.3] \\
\ln(\Lambda\Lmax') &=& -1.3(2) \quad [\gbar^2(\Lmax')=5] \, ,
\label{Lam5}
\eea
which translates into $\Lambda_{\MSbar} \Lmax = 0.36$ and
$\Lambda_{\MSbar} \Lmax' = 0.66$ with about 20\% total errors.
A corresponding number in the quenched theory \cite{CLSW} 
for $\gbar^2(\Lmax)=3.48$
is $\ln(\Lambda\Lmax)=-1.56(8)$ with the full continuum extrapolation
and $\ln(\Lambda\Lmax)=-1.47(2)$ under the present procedure with
only statistical errors indicated here.

Finally we plot the non-perturbative
evolution toward high energy for $\alpha(\mu)=\gbar^2(L)/4\pi$ ($\mu=1/L$)
starting from $\gbar^2=5$ in Fig.~5. 
\begin{figure}
  \begin{center}
    \includegraphics[width=8cm]{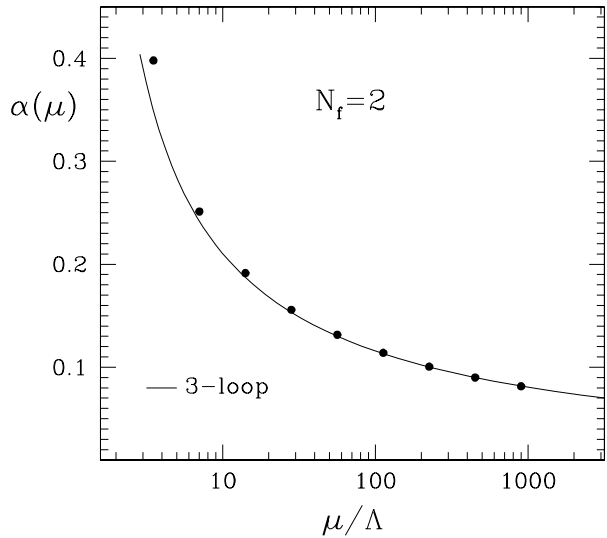}
  \end{center}
\caption{Evolution of $\alpha=\gbar^2/4\pi$ for the Schr\"odinger
functional coupling.
}
\end{figure}
Statistical errors and the difference between evolving with
$\sigma(u)$ and $\Sigmaimpr(u,1/4)$ are smaller than the symbol size.
The overall scale error 
implied by the uncertainty in the start-value
in eq.~(\ref{Lam5}),
which corresponds to a rigid horizontal shift of all data points, 
is not shown here.
In comparing the non-perturbative results with the 
    perturbative curves we emphasize that
    the important point to appreciate is that at high energies 
    the expected perturbative behaviour for our coupling has 
    been {\em shown} to set in. On the other hand the fact that the
    perturbative curves also describe the data quite well to
    rather low energies refers to a property of our particular observable
    and is definitely not to be interpreted as a reflection of some
    universal property of QCD couplings.

If, instead of evolving non-perturbatively,
we had used three-loop perturbation theory (eq.~(\ref{Lambdapert}) with $n=0$)
directly at the largest
couplings $\gbar^2=3.3 \leftrightarrow \alpha=0.26$ or
$\gbar^2=5 \leftrightarrow \alpha=0.40$,
then we would have over-estimated  $\Lambda\Lmax$ by 12\% and 23\%
respectively. This in turn translates into errors of 2\% and 5\%
for $\alpha$ in the range where its value is close to 0.12
corresponding to the physical value of $\alpha_{\MSbar}$ at $M_{\rm Z}$.
\footnote{For the estimate in the quenched approximation mentioned above the
analogous error is smaller, since there the 3-loop $\beta$-function
happens to be closer to the non-perturbative rate of evolution
over the relevant range.}

\section{Conclusions}
Our results demonstrate that with the generation of parallel computers
being installed now a computation of $\Lambda_{\MSbar}$
including two massless flavours
is becoming feasible with the ALPHA techniques. This includes
-- as in the quenched case -- the possibility to probe
and reduce systematic errors and, in particular,
lattice spacing effects. Due to the high cost of the simulations,
it is mandatory to smoothen the continuum limit as far as possible.
We have therefore spent a significant effort on accompanying perturbative 
calculations.
As observed earlier in the pure SU(2) gauge theory \cite{SU2}, we found that
lattice artefacts of several quantities constructed in the 
Schr\"odinger functional are described 
quite well by perturbation theory (see Figs.~1,2). 
This encourages us to trust in perturbation theory
to remove the lattice artefacts to a significant extent and this
procedure is supported
in the pure gauge theory in Fig.~3. As an
estimate of
the remaining systematic errors we use the difference of our results on the
finer lattices to those on the coarsest one. 
Comparing with the quenched theory, where a 
robust continuum
extrapolation could be carried out \cite{SFSU3,CLSW},
this error appears safe but also not over-pessimistic.
Nevertheless our results still need to be corroborated 
by simulations closer to the
continuum limit. We are in the process of simulating up to $L/a=16$ 
to both reduce our errors for $\Lambda L_{\rm max}$ 
and to put them on even firmer grounds, which will however still
take some time.

Already now we have clearly observed the small $N_{\rm f}$-dependence
of our discrete version of the $\beta$ function (Fig.~4).
For weak couplings its magnitude  is
accurately predicted by perturbation theory, while for our largest
coupling
($\alpha$ $\sim$ $0.25$)
it overestimates the effect significantly.
In particular, the use of perturbation theory for couplings
up to $\alpha \sim 0.4$ in estimating $\Lambda$, would lead
to a significant error already at our present level of accuracy.
Moreover, this error could hardly be quantified within the framework
of perturbation theory, which appears rather well behaved
when looked at in isolation.

Our low energy scale still has to be gauged by an experimentally
observable quantity, probably by computing
$\Lmax \fpi$. Also the extension to $N_{\rm f}=2$ of the non-perturbative 
renormalization of quark masses along the lines of ref.~\cite{CLSW}
is within reach, once the scale dependence of $\alpha$ is known.
In the more distant future we would like to
include the influence of further flavours and their masses
on the evolution of the coupling.

\vskip 1cm
\noindent
{\bf Acknowledgements.}
We thank Fred Jegerlehner for discussions. 
Our simulations have been performed on the APE computers at DESY Zeuthen.
In particular, the most compute intensive ones took advantage of APEmille.
We would like to thank DESY for early access to these machines and
the APE group in Zeuthen and in Italy,
in particular Fabio Schifano, for their continuous
most valuable  support during the early days of APEmille computing.
This work is supported in part by the 
European Community's Human Potential Programme
under contract  HPRN-CT-2000-00145
and by the Deutsche Forschungsgemeinschaft under 
Graduiertenkolleg GK 271.


\end{document}